\documentclass[a4paper]{jpconf}
\begin{document}
\title{Quantum back-reaction in a universe with positive cosmological constant}

\author{David Brizuela}

\address{
Institute for Gravitation and the Cosmos, The Pennsylvania State University,
104 Davey Lab, University Park, PA 16802, USA, and\\
Max-Planck-Institut f\"ur Gravitationsphysik, Albert-Einstein-Institut,
Am M\"uhlenberg 1, 14476 Potsdam-Golm, Germany.}

\ead{brizuela@aei.mpg.de }

\begin{abstract}
Semiclassical techniques have proven to be a very powerful method to extract
physical effects from different quantum theories. Therefore, it is
expected that in the near future they will play a very prominent role
in the context of quantum gravity. In this work we develop systematic
tools to derive semiclassical approximations for any quantum theory
with one degree of freedom. In our approach, the wave function is decomposed
in terms of an infinite set of moments, which encode the complete quantum information
of the system. Semiclassical regimes can then be properly described by truncation of this infinite
system. The use of efficient computer algebra tools allows us to compute the equations of motion
up to a very high order. In this way, we can study very precisely the quantum back reaction of the
system as well as the convergence of the method with the considered order.
Finally, these tools are applied to the particular case of a homogeneous universe
filled with a massless scalar field and positive cosmological constant, which provide
interesting physical results.
\end{abstract}

\section{Introduction}
Nowadays, one of the main goals of the research in the context of quantum gravity
is to extract physical predictions from the different theories proposed in the literature.
This is a crucial step since, in the best scenario, we will be able to make contact with
the experiments in order to verify or falsify these theories. As in any other quantum theory,
semiclassical approximations are expected to play a very prominent role in this respect.
On the one hand, they will allow us to extract new physical phenomena in their
validity regime, where quantum effects are relevant but not dominant.
On the other hand, even if deep quantum phases can be conceptually very interesting,
their direct experimental observation is not expected in a near future due to technical
difficulties. Thus, in order to obtain potentially observable effects, one should rather
focus on the semiclassical region.
In the particular case of quantum cosmology, these approximations
should provide consistency with the large-scale behavior of the universe and, probably,
some potential low energy observation, for instance through the cosmic microwave
background.

In this regard, the main available tool in quantum gravity is that of effective equations
\cite{BoSk06, BBST09}.
This approach describes the evolution
of the expectation values of the quantum observables of the system as well as its
fluctuations, correlations and higher-order moments. In the general case they form
an infinite system of couple differential equations. However in semiclassical regimes
it can be truncated and provides, to a very good approximation, the
evolution of the quantum system.

Therefore, the main goal of this work is to derive systematic semiclassical approximations
for any quantum system with one degree of freedom. In order to do so, we will first
analyze the general formalism and then construct efficient computer algebra tools.
Finally, the developed machinery will be applied to the particular case of a homogeneous
universe with a matter content of a massless scalar field and positive cosmological
constant.
 
\section{General formalism}

Let us assume we have a classical system with one degree of freedom which
is thus completely described by a couple of conjugate variables $(V,P)$.
Once the quantization of this system is performed promoting those variables
to operators, the observable information must be extracted through expectation
values. Hence, we define the infinite set of moments,
\begin{equation}\label{defG}
G^{a,b}:=\langle(\hat P-P)^a(\hat V-V)^b \rangle_{\rm Weyl},
\end{equation}
with $V:=\langle\hat V \rangle$ and $P:=\langle\hat P \rangle$,
which will encode the complete quantum information of the system.
These moments are purely quantum variables that would vanish in a classical
regime.

The evolution of these variables is then given by the effective Hamiltonian,
defined as the expectation value of the quantum Hamiltonian $\hat H$.
By Taylor expansion, it can be written as
\begin{equation}\label{Taylor}
H_{\rm eff}(V,P,G^{a,b}):=\langle\hat H\rangle_{\rm Weyl}=H(V,P) + \sum\sum\frac{1}{a!b!}
\frac{\partial^{a+b}H}{\partial P^a\partial V^b}G^{a,b},
\end{equation}
where $H$ is the classical Hamiltonian. Making use of the definition
$\{\langle\hat X\rangle,\langle\hat Y\rangle\}=-i\hbar \langle[\hat X,\hat Y]\rangle$,
for any two operators $\hat X$ and $\hat Y$, the equations of motion generated
by this effective Hamiltonian are guaranteed to be equivalent to the Schr\"odinger
flow of states or the Heisenberg flow of operators.

As has been made explicit, the effective Hamiltonian is a state-dependent object
and hence a function of expectation values and all the moments $G^{a,b}$.
Therefore, the computation of the equations of motion for different variables
implies calculating the Poisson bracket between two generic moments.
This bracket is given by,
\begin{equation}\label{GGbrackets}
\{G^{a,b},G^{c,d}\}=
a\, d \,G^{a - 1, b} \,G^{c, d - 1} - b \,c\, G^{a, b - 1}\, G^{c - 1, d}
+\sum_{n}
\left(\frac{i\hbar}{2}\right)^{n-1}
K_{abcd}^{n}\, G^{a+c-n, b+d -n},
\end{equation}
where the sum over $n$ runs over all odd numbers from 1 to
${\rm Min}(a+c,b+d,a+b,c+d)$ and the coefficients $K_{abcd}^{n}$
depend on all its indices. This formula was already presented in the literature \cite{BoSk06},
but a typo remain discovered. Due to the importance of this formula to systematize
the computation of the effective equations, in Ref. \cite{BBHKM11}, we have corrected it and
provided a complete proof.

\section{Algebraic implementation}

In order to automatize the computation of Poisson brackets between two generic
moments $\{G^{a,b},G^{c,d}\}$, we have developed
two different {\it Mathematica} codes. The first code works iteratively. That is, it begins with
the definition of the moments given by (\ref{defG}) and then uses the generic relation
$\{\langle\hat X\rangle,\langle\hat Y\rangle\}=-i\hbar \langle[\hat X,\hat Y]\rangle$,
as well as properties of the Poisson brackets and of the commutator, until all terms
are reduced to simple commutator between isolated operators $\hat V$ and $\hat P$.
This is a long process and, in fact, the time the computer requires to compute a
given Poisson bracket $\{G^{a,b},G^{c,d}\}$ by this method increases exponentially
with the considered order ($a+b+c+d$).

On the other hand, the second code makes use of the formula (\ref{GGbrackets})
and it is much more efficient. For instance, the timing to compute a 40th-order
bracket like $\{G^{11,8},G^{9,12}\}$ is less than a thousandth of a second,
which clearly shows the practical importance of equation (\ref{GGbrackets}).
This generic code is one of the main results of our research and it allows us
to obtain the equations of motion to a very high-order for any
quantum system with one degree of freedom. In the following section we will develop one example in the
context of quantum cosmology.

\section{An example: massless scalar field with positive cosmological constant}

The Friedmann equation for a homogeneous universe with a massless scalar field
$\phi$ and a positive cosmological constant $\Lambda$ reads,
\begin{equation}
 \left(\frac{a^\prime}{a}\right)^2= \frac{4\pi
 G}{3}\frac{p_{\phi}^2}{a^6}+\Lambda,
\end{equation}
where $a$ is the scale factor, $p_\phi$ the conjugate momentum of the scalar field
and derivatives are respect to cosmological time. We define the canonical gravitational
variables $V:=2 a^3/3$ and $P:=a^\prime/a$, and choose the scalar field as the
internal time. In this way, at classical level, the momentum $p_\phi$ turns out to be
the physical Hamiltonian,
\begin{equation}\label{classH}
H:= p_{\phi}= \frac{3}{2}V\sqrt{P^2-\Lambda}.
\end{equation}

In order to obtain the effective Hamiltonian that will describe the dynamics of all
moments, the Taylor expansion (\ref{Taylor}) is applied
\begin{equation}
 H_{\rm eff}:=H + \sum_{n=2}^{\infty}[V \alpha_n(P,\Lambda)G^{n,0}
+\beta_n(P, \Lambda) G^{n-1,1}].
\end{equation}
Here $\alpha_n$ and $\beta_n$ are coefficients that only depend on
the order $n$, the cosmological constant $\Lambda$ and the momentum $P$.
Note that, due to the linear dependence of the classical Hamiltonian on the
variable $V$, only moments of the form $G^{a,0}$ and $G^{a,1}$ appear on the definition
of the effective Hamiltonian.

In order to extract physical information from the system, we have obtained the
equations of motion and solve them numerically at every order up to ten.
As initial state we have chosen an unsqueezed Gaussian pulse in the volume $V$.
The corresponding moments $G^{a,b}$ are non-vanishing only if both $a$ and
$b$ are even numbers. The initial values of the expectation values $V$ and $P$,
as well as the width of the Gaussian are chosen so that initially the state is picked,
that is, the relative fluctuations $\Delta V/V$ and $\Delta P/P$ are small.
Finally, three different values have been taken for the cosmological constant:
small, intermediate and large.

The behavior of this system with vanishing cosmological constant ($\Lambda=0$)
is particularly simple. In this case, the effective Hamiltonian is only composed by a
finite amount of terms and all equations of motion decouple. Therefore there is
no back reaction and the evolution of all variables is simply exponential $V\approx e^\phi$,
$P\approx e^{-\phi}$, and $G^{a,b}\approx e^{(b-a)\phi}$. This harmonic system
is used as a reference to check that the quantum back reaction becomes stronger when
considering larger values for $\Lambda$.
On the other hand, the negative cosmological case has been analyzed with these
effective techniques in \cite{BoTa08} and with a complete numerical solution of
the wave equation in \cite{BePa08}. However, for $\Lambda<0$, the volume is bounded
by recollapse and  the moments do not grow for large volumes. These two
features are not present for $\Lambda>0$ case, making this analysis of high-order
moments necessary.

Regarding the evolution of the expectation values V and P, we would like to point
out two main features we have observed. Classically, the
volume $V(\phi)$ of the universe is an increasing function and reaches infinity at a
finite value of time $\phi=\phi_{\rm div}$. Our analysis shows that the quantum back reaction
enhances this divergence. That is, at higher order the divergence of the volume is reached faster.
The second feature is related to the
convergence. When we truncate the system at a given order, we implicitly assume
that the higher-order moments can not change significantly the observed behavior.
In order to test this convergence, we have defined $\Delta V_n(\phi):=1-V_n(\phi)/V_{n+1}(\phi)$,
where $V_n(\phi)$ is the trajectory of the expectation value of the volume at order $n$.
This object is time dependent, so its behavior will change with $\phi$.
We have studied three different values of $\phi$ and
in all the considered cases, even for very late times, and for all
values of $\Lambda$, we have obtained an exponential convergence of the system
with the order $n$. This is a very important result and shows the reliability of our
method even near the divergence of the volume.

As of the behavior of the moments, we observe that they depart from their harmonic
behavior ($\Lambda=0$ case) as the value of the cosmological constant is increased. As expected, several
moments grow and become dominant, providing thus strong quantum back reaction.
In all cases, even
for small values of $\Lambda$, as soon as the evolution is started, the initially vanishing moments
are immediately excited. In this way, the initial Gaussian shape is broken. Remarkably,
after a rapid initial increase of these moments, the state adapts itself to the evolution and moments
change less severely. One of the open questions in this respect is whether this natural
state is independent of the initial conditions. We can not give any definitive answer without
performing further numerical analysis but, due to the rapid adaptation to the evolution, 
the consequences of assuming an initial Gaussian state might, after all, be not so special.

\section{Conclusions}

We have presented a systematic method to analyze the effective behavior of any quantum
system with one degree of freedom. The key ingredients of this method are the generic
formula for the Poisson bracket between any two moments (\ref{GGbrackets}) and an
extensive use of computer algebra tools. As an example, these tools have
been applied to the particular case of a homogeneous universe with a matter content
of a massless scalar field and positive cosmological constant. The main results we have
obtained is that the quantum back reaction enhances the classical divergence of the
volume. In addition, we have observed that the initial Gaussian state is immediately
adapted to another state better suited to the evolution. And remarkably, exponential
convergence of the results with the considered order has been obtained even for late
times (near the divergence of the volume), which validates the applied methods.

\section*{Acknowledgments}
I would like to thank M. Bojowald, H.H. Hern\'andez, M. J.~Koop, and
H. A.~Morales-T\'ecotl for joint collaboration on which the presented
talk is based.
This work was supported by the Spanish
Ministry of Education through National Program No. I-D+i2008-2011
and by the Spanish MICINN Project No. FIS2008-06078-C03-03.

\end{document}